\documentclass[aps,prb,%preprint, groupedaddress
twocolumn, superscriptaddress]{revtex4-2}
\usepackage{graphicx}% Include figure files
\usepackage{booktabs}
\usepackage{hyperref}
\usepackage{soul, color, xcolor}

\begin{document}

% Use the \preprint command to place your local institutional report
% number in the upper righthand corner of the title page in preprint mode.
% Multiple \preprint commands are allowed.
% Use the 'preprintnumbers' class option to override journal defaults
% to display numbers if necessary
%\preprint{}

%Title of paper
\title{Mirror-symmetry engineering of altermagnetism based on monolayer hexagonal lattices}

% repeat the \author .. \affiliation  etc. as needed
% \email, \thanks, \homepage, \altaffiliation all apply to the current
% author. Explanatory text should go in the []'s, actual e-mail
% address or url should go in the {}'s for \email and \homepage.
% Please use the appropriate macro foreach each type of information

% \affiliation command applies to all authors since the last
% \affiliation command. The \affiliation command should follow the
% other information
% \affiliation can be followed by \email, \homepage, \thanks as well.
\author{Han-Bing Li}
\author{Zhi-Gang Shao}
\email{zgshao@scnu.edu.cn}
\affiliation{Guangdong Basic Research Center of Excellence for Structure and Fundamental Interactions of Matter, Guangdong Provincial Key Laboratory of Quantum Engineering and Quantum Materials, School of Physics, South China Normal University, Guangzhou 510006, China}
\affiliation{Guangdong-Hong Kong Joint Laboratory of Quantum Matter, Frontier Research Institute for Physics, South China Normal University, Guangzhou 510006, China}
%\email[]{Your e-mail address}
%\homepage[]{Your web page}
%\thanks{}
%\altaffiliation{}
%\affiliation{}

%Collaboration name if desired (requires use of superscriptaddress
%option in \documentclass). \noaffiliation is required (may also be
%used with the \author command).
%\collaboration can be followed by \email, \homepage, \thanks as well.
%\collaboration{}
%\noaffiliation

\date{\today}

\begin{abstract}
Altermagnetism (AM) is a novel collinear antiferromagnetic phase characterized by spin-polarized bands and zero net magnetization. However, realizing AM in monolayer hexagonal lattices is strongly constrained by lattice symmetry. Here, we propose a mirror-symmetry-based design principle and implementation scheme for realizing AM based on monolayer hexagonal lattices and validate it using holey graphyne (HGY) as a representative system. By examining hydrogenated nonmetal units X--H (X = B, C, N, P, and S) introduced along the mirror-symmetry axis and extending the analysis to N--F, we identify the configurations that form two antiparallel spin sublattices related by mirror reflection. The N--H structure forms a stable altermagnetic ground state, with a momentum-dependent spin splitting up to 148.2 meV. The magnetism originates from $\pi$--$\pi$ superexchange interactions, and H/F functionalization stabilizes intrinsic AM by regulating the local charge distribution. Our work offers insights into the design of light-element altermagnetic materials.
\end{abstract}

% insert suggested keywords - APS authors don't need to do this
%\keywords{}

%\maketitle must follow title, authors, abstract, and keywords
\maketitle

% body of paper here - Use proper section commands
% References should be done using the \cite, \ref, and \label commands
% Put \label in argument of \section for cross-referencing
%\section{\label{}}
\section{Introduction}
Altermagnetism (AM) is an unconventional collinear antiferromagnetic (AFM) phase characterized by ``spin--momentum locking''~\cite{1}. It combines the characteristics of spin-polarized bands with zero net magnetization, making it a research focus in magnetism and spintronics~\cite{1,2}. The uniqueness of altermagnets arises from opposite-spin sublattices that are related by mirror or rotational symmetry operations rather than inversion or translation~\cite{3,4}. This unique symmetry relation gives rise to momentum-dependent spin splitting in the absence of spin--orbit coupling (SOC)~\cite{5}. Consequently, altermagnets exhibit various unconventional physical phenomena, including the crystal Hall effect~\cite{6}, anomalous Hall effect~\cite{7}, giant and tunneling magnetoresistance~\cite{8}, and magneto-optical Kerr effect~\cite{9}, offering promising opportunities for applications in information storage and processing.

In two-dimensional (2D) systems, achieving controllable altermagnetic order is highly challenging due to stringent inherent symmetry constraints~\cite{7,10}. It is worth noting that the antiparallel spin sublattices in 2D altermagnets cannot be connected by mirror reflection $M_z$ or twofold rotation $C_{2z}$~\cite{11}. To overcome this limitation, researchers have attempted to break the relevant symmetries through approaches such as bilayer twisting~\cite{12}, external electric fields~\cite{13,49-fop}, and interlayer stacking~\cite{14}, thereby realizing AM~\cite{15,16,48-fop}. Currently, studies on intrinsic monolayer altermagnets mainly focus on square-lattice systems, including transition-metal compounds~\cite{17,18,19}, such as V$_2$Se$_2$O~\cite{18} and CrO~\cite{19}, as well as metal–organic frameworks (MOFs) such as Ca(pyrazine)$_2$~\cite{20} and t-Cr$_2$[Pyc-O$_8$]~\cite{21}. These works demonstrate that realizing AM in low-dimensional systems requires constructing specific symmetry mappings between inequivalent sublattices.

Compared with these square-lattice monolayer systems, studies on AM in monolayer hexagonal lattices remain at an early stage. Recent studies have shown that Janus structures can realize AM in monolayers, such as Cr$_2$Si$_2$S$_3$Se$_3$ and Mn$_2$P$_2$S$_3$Se$_3$~\cite{22,23}. In addition, ``superatom'' engineering has also been demonstrated as an effective route for designing altermagnets~\cite{24}. Its underlying mechanism is similar to that of Janus structures, both relying on intrinsic structural polarity that breaks the combined spatial-inversion and time-reversal (\textit{PT}) symmetry. To further expand the design space of 2D hexagonal altermagnets, $p$-electron systems have been considered ideal platforms for realizing nonrelativistic AM owing to their light-element composition, weak SOC, long spin lifetimes, and high tunability~\cite{25,26}. Recently, altermagnetic phases have been realized in graphene via atomic-scale hydrogen adsorption~\cite{27}. A previous study demonstrated that the adsorption of nonmagnetic $sp$ atoms at specific sites of holey graphyne (HGY) can induce altermagnetic configurations, although such configurations are not necessarily the magnetic ground state~\cite{28}. However, the underlying lattice-level design principle, its transferability to other material platforms, and a controllable route for stabilizing the non-ground-state altermagnetic configurations remain to be established.

In this Letter, we propose an altermagnetic design strategy based on monolayer hexagonal parent lattices. The central requirement is to construct two antiparallel magnetic sublattices related by a mirror operation ($M_x$ or $M_y$) rather than inversion, thereby breaking \textit{PT} symmetry and enabling momentum-dependent spin splitting. Using HGY as a representative platform, we systematically investigate the effects of introducing hydrogenated nonmetal atoms X--H (X = B, C, N, P, and S) on the magnetic ground state and electronic structure, and further examine F functionalization in the N--F system. We clarify the structural conditions required to form mirror-related spin sublattices and elucidate the microscopic mechanism underlying the emergence of AM. These results provide a symmetry-based route to realize AM in light-element monolayers. The detailed computational methods are provided in Note~S1 of the Supplemental Materials.

\section{Results and Discussion}
\subsection{Design Principle in 2D Hexagonal Lattices}

\begin{figure}[ht]
	\centering % Requires \usepackage{graphicx}
	\includegraphics[width=0.75\linewidth]{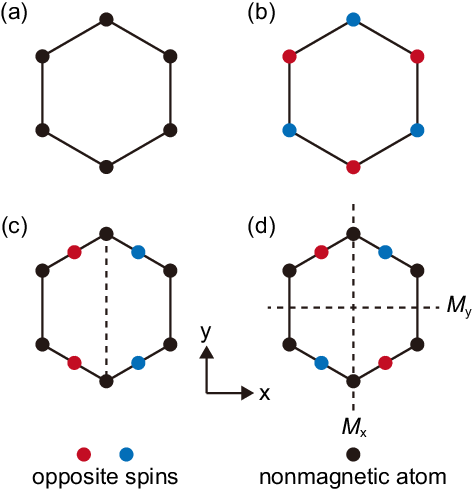}
	\caption{Design model for AM in 2D hexagonal lattices.
		(a) Schematic model of a honeycomb lattice.
		(b) Honeycomb lattice with antiparallel magnetic ordering. The lattice preserves $PT$ symmetry and exhibits a conventional AFM phase.
		(c) and (d) Configurations with nonmagnetic atoms.
		(c) The opposite-spin sublattices are related by spatial inversion, thereby preserving $PT$ symmetry.
		(d) The opposite-spin sublattices are related by $M_x$ or $M_y$ rather than by spatial inversion, thereby breaking $PT$ symmetry. The colored markers in (c) and (d) schematically represent opposite-spin sites to illustrate their symmetry relations, rather than the atomic positions of a specific material.}
	\label{Fig1}
\end{figure}

The key to constructing altermagnets lies in symmetry engineering. Figure~\ref{Fig1}(a) illustrates a 2D hexagonal honeycomb lattice. Based on this model, we introduce opposite spins to construct a lattice with antiparallel magnetic ordering. When the two opposite spins are placed on two inequivalent nonmagnetic atomic sites [Fig.~\ref{Fig1}(b)], the antiparallel spin sublattices do not possess translational symmetry $\tau$, but inversion symmetry $P$ is still preserved. As a result, the system remains invariant under the $PT$ operation, leading to spin-degenerate band structures characteristic of conventional antiferromagnets.

Typically, high crystal symmetry may be broken by introducing nonmagnetic atoms. It should be emphasized that placing antiparallel spins at centrosymmetric positions preserves the combined $PT$ symmetry, as shown in Fig.~\ref{Fig1}(c). To achieve AM, the lattice can be divided into two regions along mirror-symmetry axes $M_x$ or $M_y$, and the opposite spins can be placed such that the two sublattices are related by $M_x$ or $M_y$ rather than by spatial inversion, as illustrated in Fig.~\ref{Fig1}(d). This spin arrangement breaks $PT$ symmetry and induces altermagnetic behavior. Here, “hexagonal lattice” refers to the geometric framework of the parent monolayer. The introduction of atoms at selected sites generally removes the original threefold or sixfold rotational symmetry. Accordingly, the symmetry and wave character of the resulting altermagnetic state must be determined from the actual functionalized magnetic structure.

Furthermore, our model can be extended to the design of quasi-one-dimensional (Q1D) altermagnetic systems, as shown in Fig.~S1. By decomposing the lattice in Fig.~\ref{Fig1}(d) along $M_x$ or $M_y$, two 1D atomic chains with mirror-related antiparallel spin sites can be obtained [Figs.~S1(a)--S1(b)], which may serve as building blocks for chain-assembled Q1D monolayers. Two distinct self-assembly schemes are shown in Figs.~S1(c)--S1(d), both of which retain the mirror-related antiparallel-spin arrangement required for altermagnetic symmetry. With finite interchain coupling, these Q1D models can support momentum-dependent spin splitting. Previous studies provide evidence supporting this design strategy~\cite{38}.

\subsection{Realization of the Design Principle}

\begin{figure*}[!ht]
	\centering
	\includegraphics[width=0.9\linewidth]{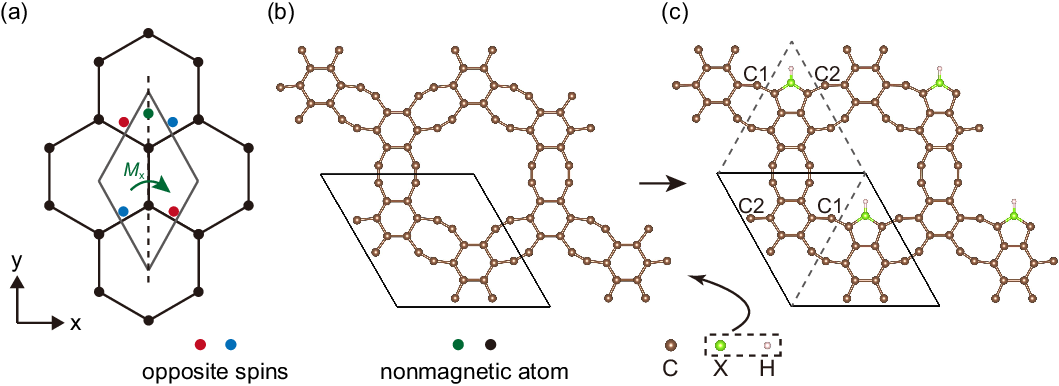}
	\caption{Implementation strategy and structural schematics for realizing the altermagnetic design principle.
		(a) Design model for achieving AM in a nonmagnetic hexagonal monolayer. By introducing nonmagnetic atoms along the symmetry axis of the unit cell, opposite spins are induced on the two sides, forming mirror-related antiparallel spin sublattices.
		(b) Geometric structure of HGY. The solid frame marks the unit cell.
		(c) Geometric structure of X--H modified HGY. The X bonds with neighboring $sp$-hybridized C atoms, breaking the acetylene bond and inducing two antiparallel magnetic carbon atoms (C1 and C2). The dashed frame represents the unit cell after rotating the lattice vectors.}
	\label{Fig2}
\end{figure*}

Based on the above design principle, we propose that AM can be realized in monolayer hexagonal lattices through symmetry engineering. For systems that intrinsically contain magnetic atoms, one may directly design mirror-related antiparallel spin sublattices according to symmetry requirements. For nonmagnetic systems, structural modulation is necessary to induce local magnetic moments that meet the symmetry requirements. As illustrated in Fig.~\ref{Fig2}(a), in a nonmagnetic monolayer, adsorbing or embedding a nonmagnetic atom along the symmetry axis of the unit cell can tune the local charge distribution and break the high crystalline symmetry, thereby inducing antiparallel magnetic atoms on both sides of the symmetry axis. The resulting two spin sublattices are related by the symmetry $[C_2 \parallel M_x]$ rather than spatial inversion, thereby breaking the combined $PT$ symmetry and giving rise to an altermagnetic phase. The key to this strategy does not lie in the specific manner of atomic introduction, but in constructing alternating antiparallel magnetic moments at mirror-symmetric positions.

The acetylenic corners between adjacent eight-membered rings in HGY provide suitable sites for inducing an altermagnetic configuration by embedding nonmetal atoms~\cite{28}. As shown in Fig.~\ref{Fig2}(b), adjacent eight-membered rings are mirror-symmetric, providing the geometric basis for realizing the model in Fig.~\ref{Fig2}(a). Importantly, impurity embedding is only one possible means of generating the required local moments rather than a necessary condition of the proposed design principle. As a distinct implementation, in the covalent organic framework COF-1~\cite{39}, removing two H atoms that saturate the electrons on both sides of the symmetry axis also generates a similar antiparallel magnetic configuration (see Note~S2), supporting the applicability of this design principle beyond HGY.

\subsection{Charge Modulation via H Functionalization for Intrinsic AM}

It should be emphasized that inducing local magnetic moments does not necessarily lead to intrinsic AM. In HGY, adsorption of B and S produces intrinsic altermagnetic ground states, whereas the altermagnetic configurations induced by N and P adsorption are not the magnetic ground states~\cite{28}. To further modulate the local charge distribution and obtain a stable altermagnetic phase, an H atom is introduced to form an X--H pair, as illustrated in Figs.~\ref{Fig2}(b)--\ref{Fig2}(c). We systematically examined five X--H embedded systems (X = B, C, N, P, and S). In the B–H system, formation of the B–H bond modifies the local coordination and $p$-orbital hybridization of B, driving the B–H unit and the neighboring carbon framework out of the HGY plane. The coupled structural reconstruction and charge redistribution leave no unpaired electrons on C1 and C2, resulting in a nonmagnetic state. In the S–H system, the S atom favors a two-coordinate local environment in the present HGY framework. Consequently, it bonds with only one neighboring C atom and cannot generate the required mirror-related magnetic sites. Therefore, neither system exhibits altermagnetic behavior. In contrast, the C--H, N--H, and P--H configurations preserve the required $M_x$ symmetry and spin distribution of the model in Fig.~\ref{Fig2}(a). They induce local magnetic moments on C1 and C2 on opposite sides of the symmetry axis. Within the unit cell, the considered AFM configurations are lower in energy than the corresponding ferromagnetic (FM) configurations by 28.2 meV (C--H), 58.5 meV (N--H), and 164.9 meV (P--H), respectively. For comparison, in C-embedded HGY, the embedded C atom carries a finite magnetic moment, and the parallel configuration is energetically favored over the antiparallel configuration. This contrast indicates that H functionalization modifies the local spin distribution and the preferred alignment of the C1 and C2 moments.

\begin{figure*}[!ht]
	\centering
	\includegraphics[width=0.9\linewidth]{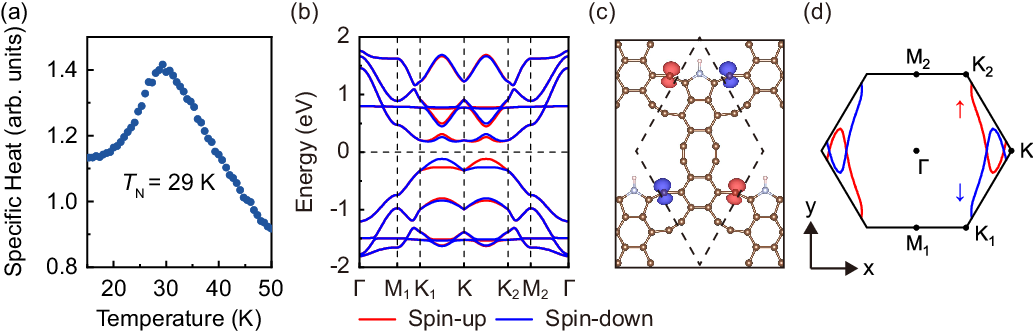}
	\caption{(a) Temperature-dependent specific heat of the N--H system obtained from the anisotropic Heisenberg model. The specific-heat anomaly occurs at approximately 29~K.
		(b) Non-relativistic spin-polarized band structure of the N--H system.
		(c) Spin density distribution of the N--H system. The dashed frame indicates the unit cell after rotating the lattice vectors. Red and blue denote spin-up and spin-down densities, respectively. The two antiparallel spin sublattices are connected by the combined operation $[C_2\Vert M_x]$.
		(d) Constant-energy contours of the valence bands near the Fermi level. Red and blue arrows denote spin-up and spin-down channels, respectively.}
	\label{Fig3}
\end{figure*}

To evaluate the structural stability of the C--H, N--H, and P--H systems, \textit{ab initio} molecular dynamics simulations were performed, as shown in Fig.~S3. At 300\,K, both the C--H and N--H structures remain intact, with their total energies fluctuating slightly around constant values, indicating good thermal stability. In contrast, the P--H system exhibits bond breaking and pronounced energy fluctuations, reflecting its poor thermal stability. Therefore, the P--H system is excluded from further consideration. Further phonon calculations reveal no apparent imaginary frequencies in either the C--H or N--H system along the sampled high-symmetry path (Fig.~S4), confirming their dynamical stability.
	
To determine the magnetic ground states of the C--H and N--H systems, we compared four magnetic configurations (AFM-1, AFM-2, AFM-3, and FM). The corresponding spin arrangements are illustrated using the N--H system as a representative structure in Fig.~S5. For C--H, AFM-2 has the lowest total energy, whereas the altermagnetic configuration corresponds to AFM-1. The exchange parameters are $J_1=-15.7$ meV, $J_2=1.2$ meV, and $J_3=-0.3$ meV. The positive $J_2$ indicates FM second-nearest-neighbor coupling, which prevents AM from becoming the magnetic ground state. The C--H unit and a single N atom both nominally contribute five valence electrons, consistent with the similar magnetic ordering obtained for C--H-embedded and N-adsorbed HGY~\cite{28}.

In contrast, AFM-1 is the lowest-energy configuration of the N--H system. Its exchange parameters, $J_1=-21.4$ meV, $J_2=-6.4$ meV, and $J_3=-1.3$ meV, are all negative, indicating AFM coupling and stabilizing AFM-1 as the intrinsic altermagnetic ground state. Compared with N-adsorbed HGY, H functionalization changes the second-nearest-neighbor interaction from FM to AFM, demonstrating its role in regulating the exchange interactions and magnetic ground state. Therefore, subsequent analyses focus on the N--H system. SOC calculations identify the $x$ direction as the easy axis, with its energy lower than those of the $y$ and $z$ orientations by 41.1 and 31.3~$\mu$eV per unit cell, respectively. Monte Carlo simulations based on the anisotropic Heisenberg model yield an estimated N\'eel temperature of approximately 29~K [Fig.~\ref{Fig3}(a)]. Further details are provided in Note~S3.

Experimentally, the proposed N--H/HGY system may be pursued by first constructing the HGY framework through bottom-up assembly, followed by the controlled formation of N--H-containing bridging motifs at the acetylenic corner sites. The experimental construction of in-plane B-based bridges in 2D $g$-C$_3$N$_4$ through supercritical-CO$_2$-assisted treatment provides a relevant precedent for this bridge-functionalization strategy~\cite{47}. 

The band structure of the N--H system indicates that it exhibits semiconducting behavior with a band gap of 0.3~eV, as shown in Fig.~\ref{Fig3}(b). This can be attributed to the even number of electrons introduced by the N and H atoms. Notably, a pronounced spin splitting occurs along the K$_1$--K--K$_2$ path. Importantly, this spin splitting is not induced by SOC, further confirming the presence of AM, consistent with the symmetry analysis. As shown in Fig.~\ref{Fig3}(c), the magnetic moments of the N--H system primarily originate from C1 and C2, which possess opposite spins and are related by the $M_x$ symmetry, leading to the breaking of $PT$ symmetry. In the collinear spin-polarized calculations, C1 and C2 have projected magnetic moments of +0.512 and -0.512 $\mu_{\mathrm B}$ along the same spin-quantization axis, respectively. Thus, their moments are collinear and antiparallel, and the total magnetic moment is 0 $\mu_{\mathrm B}$ per unit cell. Moreover, the spin-split bands on the two sides of the K point exhibit opposite spin polarizations. Under the combined operation $[C_2\Vert M_x]$, the two spin channels satisfy $E_{\uparrow}(k_x,k_y)=E_{\downarrow}(-k_x,k_y)$. Together with the constant-energy contours in Fig.~\ref{Fig3}(d), this behavior identifies a $d$-wave-type altermagnetic spin-splitting pattern. It differs from the $i$-wave pattern expected when the relevant threefold/sixfold symmetry of a hexagonal magnetic crystal is retained. The spin splitting in the N--H system reaches 148.2~meV, which is smaller than that of N-adsorbed HGY without H functionalization (205.1~meV)~\cite{28}, but remains larger than those reported for several other 2D altermagnets, including the Cr-based MOFs Cr(diz)$_2$, Cr($c$-pyr)$_2$, and Cr($f$-pid)$_2$ (16.48–37.54 meV)~\cite{41} and MnP(S,Se)$_3$ ($\sim$100~meV)~\cite{23}.

\begin{figure*}[!t]
	\centering
	\includegraphics[width=0.8\linewidth]{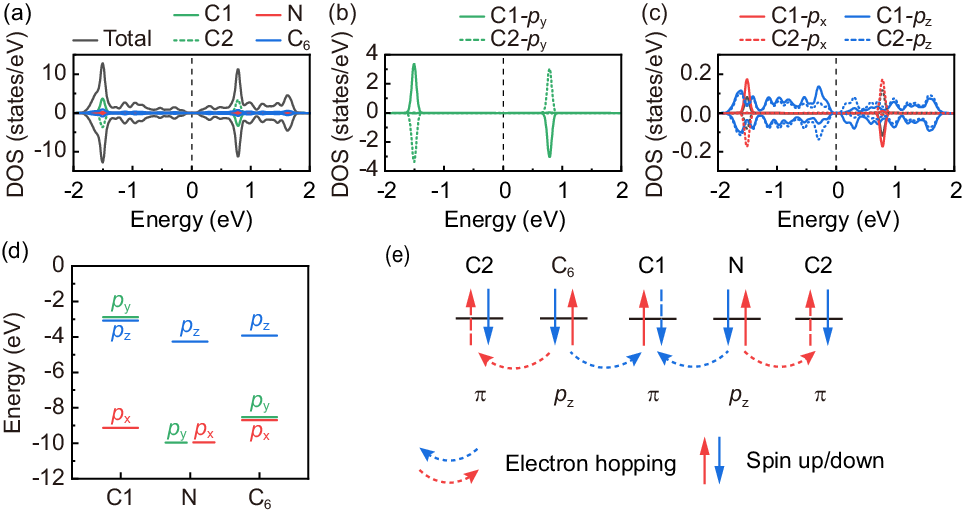}
	\caption{(a) Total and atom-projected DOS of the N--H system.
		(b) and (c) Orbital-resolved DOS of the magnetic atoms C1 (solid line) and C2 (dashed line).
		(d) Orbital-level alignment among C1, N, and the carbon ring (C$_6$).
		(e) $\pi$--$\pi$ superexchange pathway between two neighboring magnetic C atoms. C$_6$ and N act as mediators in the superexchange process, with occupied and spin-compensated $p_z$ orbitals, while C1 and C2 possess spin-polarized $\pi$ orbitals. Only the partially occupied $\pi$ orbitals are shown.}
	\label{Fig4}
\end{figure*}

To quantify the charge modulation induced by H, Bader charge analysis was performed for the N–H system. Approximately 0.684 $e$ is transferred from the HGY framework to the N–H group. At the atomic level, the N atom gains 1.156 $e$, whereas the H atom loses 0.472 $e$, revealing pronounced internal charge polarization from H toward N. These results provide direct quantitative evidence that H functionalization regulates the local charge distribution in the N–H system.

We further examined three representative stacking arrangements of N--H-embedded HGY bilayers while retaining the intralayer altermagnetic spin pattern. Among the six configurations considered, the half-lattice-shifted S1/2-AFM configuration has the lowest energy and retains altermagnetic characteristics, supporting extension beyond the monolayer. Details are provided in Note~S4.

\subsection{Microscopic Origin of the Antiparallel Magnetic Ordering}

As shown in Fig.~\ref{Fig4}(a), the density of states (DOS) of C1 and C2 is spin-asymmetric, exhibiting pronounced spin polarization and serving as the primary source of magnetism in the system. Due to the compensating contributions of C1 and C2, the total DOS is fully symmetric, resulting in zero net magnetization. The orbital-resolved DOS in Figs.~\ref{Fig4}(b)--\ref{Fig4}(c) further reveals that the magnetism mainly arises from the $p_y$ orbitals of C1 and C2, with minor contributions from the $p_x$ and $p_z$ orbitals. In contrast, the orbital-resolved DOS of the intermediate N atom and the carbon ring C$_6$ are completely symmetric (Fig.~S8), indicating zero net magnetic moment. This conclusion is further supported by the orbital-level (global orbital centers) alignment shown in Fig.~S9. For C1, all spin-up channels of the $p$ orbitals lie at lower energies, giving rise to a positive magnetic moment. Among them, the $p_y$ orbital exhibits the most pronounced spin splitting, showing the largest difference in electron occupancy between spin channels, followed by $p_z$, while $p_x$ shows the smallest splitting. Therefore, the unpaired electrons predominantly occupy the spin-up $p_y$ and $p_z$ orbitals. Conversely, C2 exhibits the opposite spin polarization. All spin-down channels of the $p$ orbitals are located at lower energies than the corresponding spin-up channels, leading to an identical but opposite negative magnetic moment. In comparison, all orbitals of N and C$_6$ exhibit nearly identical energy levels and electron occupancies in the two spin channels, with the occupied and spin-compensated $p$ orbitals.

To reveal the origin of AM in the N--H system, it is essential to determine the mechanism for the antiparallel alignment of the magnetic atoms. Figure~S10 shows the occupied-state orbital energy levels ($\varepsilon_{\rm occ}$) of C1, N, and C$_6$, defined as
\[
\varepsilon_{\rm occ} = \frac{\int_{-\infty}^{E_{\mathrm{F}}} E \, g(E) \, dE}{\int_{-\infty}^{E_{\mathrm{F}}} g(E) \, dE},
\]
where $g(E)$ is the DOS and $E_{\mathrm{F}}$ is the Fermi level. As shown in Fig.~\ref{Fig4}(d), the $p_y$ and $p_z$ orbitals of C1 form high-energy $\pi$ states. Their energies are close to those of the $p_z$ orbitals of N and C$_6$, supporting the $\pi$--$\pi$ superexchange pathway illustrated in Fig.~\ref{Fig4}(e). The spin-polarized $\pi$ orbitals of C1/C2 interact with the occupied and spin-compensated $p_z$ orbitals of N and C$_6$ via virtual electron hopping, resulting in superexchange interactions. Since C1/C2, N, and C$_6$ are nearly collinear, according to the Goodenough--Kanamori--Anderson rules~\cite{42,43,44}, this favors antiparallel spin alignment between C1 and C2, thereby stabilizing the altermagnetic ground state of the N--H system.

\subsection{Extension to N--F Functionalization}
The above results suggest that the essential role of H functionalization lies in regulating the local charge distribution and orbital occupation, rather than in introducing an additional magnetic center. Previous studies have explored the extension of H-functionalized 2D materials to their halogen-functionalized counterparts~\cite{45,46}. Inspired by this strategy, we further considered F functionalization, namely N--F embedded in a HGY monolayer [Fig.~S11(a)]. The N--F system exhibits electronic and magnetic characteristics similar to those of the N--H system and also possesses an altermagnetic ground state. The projected magnetic moments on C1 and C2 are +0.515 and -0.515 $\mu_{\mathrm B}$, respectively, while the total magnetic moment is 0 $\mu_{\mathrm B}$ per unit cell. SOC calculations indicate an effective $xz$ easy plane, while anisotropic Heisenberg Monte Carlo simulations yield a specific-heat anomaly at approximately 26~K. Detailed discussion is provided in Note~S6.

\section{Conclusions}
In summary, we propose an AM design principle applicable to 2D hexagonal lattice systems. By constructing antiparallel spin sublattices that are mutually mapped through mirror symmetry, we achieve $PT$-symmetry breaking and momentum-dependent spin splitting in the absence of SOC. This principle is realized in HGY and further validated in COF-1, providing initial evidence of its transferability beyond a single material platform. Based on this principle, we introduce nonmetal atoms with H/F functionalization into the HGY platform. The results indicate that H/F functionalization enables a transition from a non-intrinsic to an intrinsic altermagnetic state by modulating the local charge distribution. In particular, the N--H system forms a stable altermagnetic ground state, exhibiting a $T_{\mathrm{N}}$ of 29~K and a momentum-dependent spin splitting of 148.2~meV, with the magnetism originating from $\pi$--$\pi$ superexchange interactions between \textit{sp}-hybridized carbon atoms. This work establishes a design strategy in which mirror-symmetry engineering combined with charge modulation drives the formation of AM, and verifies its feasibility in monolayer light-element hexagonal lattice materials.

\begin{acknowledgments}
	This work was supported by the National Natural Science Foundation of China (Grant No. 52072132).
\end{acknowledgments}


\begin{thebibliography}{39}%
	\makeatletter
	\providecommand \@ifxundefined [1]{%
		\@ifx{#1\undefined}
	}%
	\providecommand \@ifnum [1]{%
		\ifnum #1\expandafter \@firstoftwo
		\else \expandafter \@secondoftwo
		\fi
	}%
	\providecommand \@ifx [1]{%
		\ifx #1\expandafter \@firstoftwo
		\else \expandafter \@secondoftwo
		\fi
	}%
	\providecommand \natexlab [1]{#1}%
	\providecommand \enquote  [1]{``#1''}%
	\providecommand \bibnamefont  [1]{#1}%
	\providecommand \bibfnamefont [1]{#1}%
	\providecommand \citenamefont [1]{#1}%
	\providecommand \href@noop [0]{\@secondoftwo}%
	\providecommand \href [0]{\begingroup \@sanitize@url \@href}%
	\providecommand \@href[1]{\@@startlink{#1}\@@href}%
	\providecommand \@@href[1]{\endgroup#1\@@endlink}%
	\providecommand \@sanitize@url [0]{\catcode `\\12\catcode `\$12\catcode
		`\&12\catcode `\#12\catcode `\^12\catcode `\_12\catcode `\%12\relax}%
	\providecommand \@@startlink[1]{}%
	\providecommand \@@endlink[0]{}%
	\providecommand \url  [0]{\begingroup\@sanitize@url \@url }%
	\providecommand \@url [1]{\endgroup\@href {#1}{\urlprefix }}%
	\providecommand \urlprefix  [0]{URL }%
	\providecommand \Eprint [0]{\href }%
	\providecommand \doibase [0]{https://doi.org/}%
	\providecommand \selectlanguage [0]{\@gobble}%
	\providecommand \bibinfo  [0]{\@secondoftwo}%
	\providecommand \bibfield  [0]{\@secondoftwo}%
	\providecommand \translation [1]{[#1]}%
	\providecommand \BibitemOpen [0]{}%
	\providecommand \bibitemStop [0]{}%
	\providecommand \bibitemNoStop [0]{.\EOS\space}%
	\providecommand \EOS [0]{\spacefactor3000\relax}%
	\providecommand \BibitemShut  [1]{\csname bibitem#1\endcsname}%
	\let\auto@bib@innerbib\@empty
	%</preamble>
	\bibitem [{\citenamefont {\ifmmode~\check{S}\else \v{S}\fi{}mejkal}\ \emph
		{et~al.}(2022{\natexlab{a}})\citenamefont {\ifmmode~\check{S}\else
			\v{S}\fi{}mejkal}, \citenamefont {Sinova},\ and\ \citenamefont
		{Jungwirth}}]{1}%
	\BibitemOpen
	\bibfield  {author} {\bibinfo {author} {\bibfnamefont {L.}~\bibnamefont
			{\ifmmode~\check{S}\else \v{S}\fi{}mejkal}}, \bibinfo {author} {\bibfnamefont
			{J.}~\bibnamefont {Sinova}},\ and\ \bibinfo {author} {\bibfnamefont
			{T.}~\bibnamefont {Jungwirth}},\ }\bibfield  {title} {\bibinfo {title}
		{{Beyond Conventional Ferromagnetism and Antiferromagnetism: A Phase with
				Nonrelativistic Spin and Crystal Rotation Symmetry}},\ }\href
	{https://doi.org/10.1103/PhysRevX.12.031042} {\bibfield  {journal} {\bibinfo
			{journal} {Phys. Rev. X}\ }\textbf {\bibinfo {volume} {12}},\ \bibinfo
		{pages} {031042} (\bibinfo {year} {2022}{\natexlab{a}})}\BibitemShut
	{NoStop}%
	\bibitem [{\citenamefont {Bai}\ \emph {et~al.}(2024)\citenamefont {Bai},
		\citenamefont {Feng}, \citenamefont {Liu}, \citenamefont {{\v{S}}mejkal},
		\citenamefont {Mokrousov},\ and\ \citenamefont {Yao}}]{2}%
	\BibitemOpen
	\bibfield  {author} {\bibinfo {author} {\bibfnamefont {L.}~\bibnamefont
			{Bai}}, \bibinfo {author} {\bibfnamefont {W.}~\bibnamefont {Feng}}, \bibinfo
		{author} {\bibfnamefont {S.}~\bibnamefont {Liu}}, \bibinfo {author}
		{\bibfnamefont {L.}~\bibnamefont {{\v{S}}mejkal}}, \bibinfo {author}
		{\bibfnamefont {Y.}~\bibnamefont {Mokrousov}},\ and\ \bibinfo {author}
		{\bibfnamefont {Y.}~\bibnamefont {Yao}},\ }\bibfield  {title} {\bibinfo
		{title} {{Altermagnetism: Exploring New Frontiers in Magnetism and
				Spintronics}},\ }\href
	{https://doi.org/https://doi.org/10.1002/adfm.202409327} {\bibfield
		{journal} {\bibinfo  {journal} {Adv. Funct. Mater.}\ }\textbf {\bibinfo
			{volume} {34}},\ \bibinfo {pages} {2409327} (\bibinfo {year}
		{2024})}\BibitemShut {NoStop}%
	\bibitem [{\citenamefont {Song}\ \emph {et~al.}(2025)\citenamefont {Song},
		\citenamefont {Bai}, \citenamefont {Zhou}, \citenamefont {Han}, \citenamefont
		{Reichlova}, \citenamefont {Dil}, \citenamefont {Liu}, \citenamefont {Chen},\
		and\ \citenamefont {Pan}}]{3}%
	\BibitemOpen
	\bibfield  {author} {\bibinfo {author} {\bibfnamefont {C.}~\bibnamefont
			{Song}}, \bibinfo {author} {\bibfnamefont {H.}~\bibnamefont {Bai}}, \bibinfo
		{author} {\bibfnamefont {Z.}~\bibnamefont {Zhou}}, \bibinfo {author}
		{\bibfnamefont {L.}~\bibnamefont {Han}}, \bibinfo {author} {\bibfnamefont
			{H.}~\bibnamefont {Reichlova}}, \bibinfo {author} {\bibfnamefont {J.~H.}\
			\bibnamefont {Dil}}, \bibinfo {author} {\bibfnamefont {J.}~\bibnamefont
			{Liu}}, \bibinfo {author} {\bibfnamefont {X.}~\bibnamefont {Chen}},\ and\
		\bibinfo {author} {\bibfnamefont {F.}~\bibnamefont {Pan}},\ }\bibfield
	{title} {\bibinfo {title} {{Altermagnets as a new class of functional
				materials}},\ }\href {https://doi.org/10.1038/s41578-025-00779-1} {\bibfield
		{journal} {\bibinfo  {journal} {Nat. Rev. Mater.}\ }\textbf {\bibinfo
			{volume} {10}},\ \bibinfo {pages} {473} (\bibinfo {year} {2025})}\BibitemShut
	{NoStop}%
	\bibitem [{\citenamefont {\ifmmode~\check{S}\else \v{S}\fi{}mejkal}\ \emph
		{et~al.}(2022{\natexlab{b}})\citenamefont {\ifmmode~\check{S}\else
			\v{S}\fi{}mejkal}, \citenamefont {Sinova},\ and\ \citenamefont
		{Jungwirth}}]{4}%
	\BibitemOpen
	\bibfield  {author} {\bibinfo {author} {\bibfnamefont {L.}~\bibnamefont
			{\ifmmode~\check{S}\else \v{S}\fi{}mejkal}}, \bibinfo {author} {\bibfnamefont
			{J.}~\bibnamefont {Sinova}},\ and\ \bibinfo {author} {\bibfnamefont
			{T.}~\bibnamefont {Jungwirth}},\ }\bibfield  {title} {\bibinfo {title}
		{{Emerging Research Landscape of Altermagnetism}},\ }\href
	{https://doi.org/10.1103/PhysRevX.12.040501} {\bibfield  {journal} {\bibinfo
			{journal} {Phys. Rev. X}\ }\textbf {\bibinfo {volume} {12}},\ \bibinfo
		{pages} {040501} (\bibinfo {year} {2022}{\natexlab{b}})}\BibitemShut
	{NoStop}%
	\bibitem [{\citenamefont {Hayami}\ \emph {et~al.}(2019)\citenamefont {Hayami},
		\citenamefont {Yanagi},\ and\ \citenamefont {Kusunose}}]{5}%
	\BibitemOpen
	\bibfield  {author} {\bibinfo {author} {\bibfnamefont {S.}~\bibnamefont
			{Hayami}}, \bibinfo {author} {\bibfnamefont {Y.}~\bibnamefont {Yanagi}},\
		and\ \bibinfo {author} {\bibfnamefont {H.}~\bibnamefont {Kusunose}},\
	}\bibfield  {title} {\bibinfo {title} {{Momentum-Dependent Spin Splitting by
				Collinear Antiferromagnetic Ordering}},\ }\href
	{https://doi.org/10.7566/JPSJ.88.123702} {\bibfield  {journal} {\bibinfo
			{journal} {J. Phys. Soc. Jpn.}\ }\textbf {\bibinfo {volume} {88}},\ \bibinfo
		{pages} {123702} (\bibinfo {year} {2019})}\BibitemShut {NoStop}%
	\bibitem [{\citenamefont {\v{S}mejkal}\ \emph {et~al.}(2020)\citenamefont
		{\v{S}mejkal}, \citenamefont {Gonz\'{a}lez-Hern\'{a}ndez}, \citenamefont
		{Jungwirth},\ and\ \citenamefont {Sinova}}]{6}%
	\BibitemOpen
	\bibfield  {author} {\bibinfo {author} {\bibfnamefont {L.}~\bibnamefont
			{\v{S}mejkal}}, \bibinfo {author} {\bibfnamefont {R.}~\bibnamefont
			{Gonz\'{a}lez-Hern\'{a}ndez}}, \bibinfo {author} {\bibfnamefont
			{T.}~\bibnamefont {Jungwirth}},\ and\ \bibinfo {author} {\bibfnamefont
			{J.}~\bibnamefont {Sinova}},\ }\bibfield  {title} {\bibinfo {title} {{Crystal
				time-reversal symmetry breaking and spontaneous Hall effect in collinear
				antiferromagnets}},\ }\href {https://doi.org/10.1126/sciadv.aaz8809}
	{\bibfield  {journal} {\bibinfo  {journal} {Sci. Adv.}\ }\textbf {\bibinfo
			{volume} {6}},\ \bibinfo {pages} {eaaz8809} (\bibinfo {year}
		{2020})}\BibitemShut {NoStop}%
	\bibitem [{\citenamefont {Sheoran}\ and\ \citenamefont {Dev}(2025)}]{7}%
	\BibitemOpen
	\bibfield  {author} {\bibinfo {author} {\bibfnamefont {S.}~\bibnamefont
			{Sheoran}}\ and\ \bibinfo {author} {\bibfnamefont {P.}~\bibnamefont {Dev}},\
	}\bibfield  {title} {\bibinfo {title} {{Spontaneous anomalous Hall effect in
				two-dimensional altermagnets}},\ }\href
	{https://doi.org/10.1103/PhysRevB.111.184407} {\bibfield  {journal} {\bibinfo
			{journal} {Phys. Rev. B}\ }\textbf {\bibinfo {volume} {111}},\ \bibinfo
		{pages} {184407} (\bibinfo {year} {2025})}\BibitemShut {NoStop}%
	\bibitem [{\citenamefont {\ifmmode~\check{S}\else \v{S}\fi{}mejkal}\ \emph
		{et~al.}(2022{\natexlab{c}})\citenamefont {\ifmmode~\check{S}\else
			\v{S}\fi{}mejkal}, \citenamefont {Hellenes}, \citenamefont
		{Gonz\'alez-Hern\'andez}, \citenamefont {Sinova},\ and\ \citenamefont
		{Jungwirth}}]{8}%
	\BibitemOpen
	\bibfield  {author} {\bibinfo {author} {\bibfnamefont {L.}~\bibnamefont
			{\ifmmode~\check{S}\else \v{S}\fi{}mejkal}}, \bibinfo {author} {\bibfnamefont
			{A.~B.}\ \bibnamefont {Hellenes}}, \bibinfo {author} {\bibfnamefont
			{R.}~\bibnamefont {Gonz\'alez-Hern\'andez}}, \bibinfo {author} {\bibfnamefont
			{J.}~\bibnamefont {Sinova}},\ and\ \bibinfo {author} {\bibfnamefont
			{T.}~\bibnamefont {Jungwirth}},\ }\bibfield  {title} {\bibinfo {title}
		{{Giant and Tunneling Magnetoresistance in Unconventional Collinear
				Antiferromagnets with Nonrelativistic Spin-Momentum Coupling}},\ }\href
	{https://doi.org/10.1103/PhysRevX.12.011028} {\bibfield  {journal} {\bibinfo
			{journal} {Phys. Rev. X}\ }\textbf {\bibinfo {volume} {12}},\ \bibinfo
		{pages} {011028} (\bibinfo {year} {2022}{\natexlab{c}})}\BibitemShut
	{NoStop}%
	\bibitem [{\citenamefont {Sun}\ \emph {et~al.}(2025)\citenamefont {Sun},
		\citenamefont {Wang}, \citenamefont {Yang}, \citenamefont {Huang},
		\citenamefont {Ding}, \citenamefont {Dong},\ and\ \citenamefont {Cheng}}]{9}%
	\BibitemOpen
	\bibfield  {author} {\bibinfo {author} {\bibfnamefont {W.}~\bibnamefont
			{Sun}}, \bibinfo {author} {\bibfnamefont {W.}~\bibnamefont {Wang}}, \bibinfo
		{author} {\bibfnamefont {C.}~\bibnamefont {Yang}}, \bibinfo {author}
		{\bibfnamefont {S.}~\bibnamefont {Huang}}, \bibinfo {author} {\bibfnamefont
			{N.}~\bibnamefont {Ding}}, \bibinfo {author} {\bibfnamefont {S.}~\bibnamefont
			{Dong}},\ and\ \bibinfo {author} {\bibfnamefont {Z.}~\bibnamefont {Cheng}},\
	}\bibfield  {title} {\bibinfo {title} {{Designing Spin Symmetry for
				Altermagnetism with Strong Magnetoelectric Coupling}},\ }\href
	{https://doi.org/https://doi.org/10.1002/advs.202503235} {\bibfield
		{journal} {\bibinfo  {journal} {Adv. Sci.}\ }\textbf {\bibinfo {volume}
			{12}},\ \bibinfo {pages} {e03235} (\bibinfo {year} {2025})}\BibitemShut
	{NoStop}%
	\bibitem [{\citenamefont {Gonz{\'a}lez}\ \emph {et~al.}(2025)\citenamefont
		{Gonz{\'a}lez}, \citenamefont {Le{\'o}n}, \citenamefont
		{Gonz{\'a}lez-Fuentes},\ and\ \citenamefont {Gallardo}}]{10}%
	\BibitemOpen
	\bibfield  {author} {\bibinfo {author} {\bibfnamefont {J.~W.}\ \bibnamefont
			{Gonz{\'a}lez}}, \bibinfo {author} {\bibfnamefont {A.~M.}\ \bibnamefont
			{Le{\'o}n}}, \bibinfo {author} {\bibfnamefont {C.}~\bibnamefont
			{Gonz{\'a}lez-Fuentes}},\ and\ \bibinfo {author} {\bibfnamefont {R.~A.}\
			\bibnamefont {Gallardo}},\ }\bibfield  {title} {\bibinfo {title}
		{Altermagnetism in two-dimensional {Ca}$_2${RuO}$_4$ perovskite},\ }\href
	{https://doi.org/10.1039/D4NR04053H} {\bibfield  {journal} {\bibinfo
			{journal} {Nanoscale}\ }\textbf {\bibinfo {volume} {17}},\ \bibinfo {pages}
		{4796} (\bibinfo {year} {2025})}\BibitemShut {NoStop}%
	\bibitem [{\citenamefont {Zeng}\ and\ \citenamefont {Zhao}(2024)}]{11}%
	\BibitemOpen
	\bibfield  {author} {\bibinfo {author} {\bibfnamefont {S.}~\bibnamefont
			{Zeng}}\ and\ \bibinfo {author} {\bibfnamefont {Y.-J.}\ \bibnamefont
			{Zhao}},\ }\bibfield  {title} {\bibinfo {title} {{Description of
				two-dimensional altermagnetism: Categorization using spin group theory}},\
	}\href {https://doi.org/10.1103/PhysRevB.110.054406} {\bibfield  {journal}
		{\bibinfo  {journal} {Phys. Rev. B}\ }\textbf {\bibinfo {volume} {110}},\
		\bibinfo {pages} {054406} (\bibinfo {year} {2024})}\BibitemShut {NoStop}%
	\bibitem [{\citenamefont {Liu}\ \emph {et~al.}(2024)\citenamefont {Liu},
		\citenamefont {Yu},\ and\ \citenamefont {Liu}}]{12}%
	\BibitemOpen
	\bibfield  {author} {\bibinfo {author} {\bibfnamefont {Y.}~\bibnamefont
			{Liu}}, \bibinfo {author} {\bibfnamefont {J.}~\bibnamefont {Yu}},\ and\
		\bibinfo {author} {\bibfnamefont {C.-C.}\ \bibnamefont {Liu}},\ }\bibfield
	{title} {\bibinfo {title} {{Twisted Magnetic Van der Waals Bilayers: An Ideal
				Platform for Altermagnetism}},\ }\href
	{https://doi.org/10.1103/PhysRevLett.133.206702} {\bibfield  {journal}
		{\bibinfo  {journal} {Phys. Rev. Lett.}\ }\textbf {\bibinfo {volume} {133}},\
		\bibinfo {pages} {206702} (\bibinfo {year} {2024})}\BibitemShut {NoStop}%
	\bibitem [{\citenamefont {Wang}\ \emph {et~al.}(2025)\citenamefont {Wang},
		\citenamefont {Wang}, \citenamefont {Liu}, \citenamefont {Zhang},\ and\
		\citenamefont {Zhang}}]{13}%
	\BibitemOpen
	\bibfield  {author} {\bibinfo {author} {\bibfnamefont {D.}~\bibnamefont
			{Wang}}, \bibinfo {author} {\bibfnamefont {H.}~\bibnamefont {Wang}}, \bibinfo
		{author} {\bibfnamefont {L.}~\bibnamefont {Liu}}, \bibinfo {author}
		{\bibfnamefont {J.}~\bibnamefont {Zhang}},\ and\ \bibinfo {author}
		{\bibfnamefont {H.}~\bibnamefont {Zhang}},\ }\bibfield  {title} {\bibinfo
		{title} {{Electric-Field-Induced Switchable Two-Dimensional Altermagnets}},\
	}\href {https://doi.org/10.1021/acs.nanolett.4c05384} {\bibfield  {journal}
		{\bibinfo  {journal} {Nano Lett.}\ }\textbf {\bibinfo {volume} {25}},\
		\bibinfo {pages} {498} (\bibinfo {year} {2025})}\BibitemShut {NoStop}%
	\bibitem [{\citenamefont {Guo}(2026)}]{49-fop}%
	\BibitemOpen
	\bibfield  {author} {\bibinfo {author} {\bibfnamefont {S.-D.}\ \bibnamefont
			{Guo}},\ }\bibfield  {title} {\bibinfo {title} {{Hidden altermagnetism}},\
	}\href {https://doi.org/https://doi.org/10.15302/frontphys.2026.025201}
	{\bibfield  {journal} {\bibinfo  {journal} {Front. Phys.}\ }\textbf {\bibinfo
			{volume} {21}},\ \bibinfo {pages} {025201} (\bibinfo {year}
		{2026})}\BibitemShut {NoStop}%
	\bibitem [{\citenamefont {Pan}\ \emph {et~al.}(2024)\citenamefont {Pan},
		\citenamefont {Zhou}, \citenamefont {Lyu}, \citenamefont {Xiao},
		\citenamefont {Yang},\ and\ \citenamefont {Sun}}]{14}%
	\BibitemOpen
	\bibfield  {author} {\bibinfo {author} {\bibfnamefont {B.}~\bibnamefont
			{Pan}}, \bibinfo {author} {\bibfnamefont {P.}~\bibnamefont {Zhou}}, \bibinfo
		{author} {\bibfnamefont {P.}~\bibnamefont {Lyu}}, \bibinfo {author}
		{\bibfnamefont {H.}~\bibnamefont {Xiao}}, \bibinfo {author} {\bibfnamefont
			{X.}~\bibnamefont {Yang}},\ and\ \bibinfo {author} {\bibfnamefont
			{L.}~\bibnamefont {Sun}},\ }\bibfield  {title} {\bibinfo {title} {{General
				Stacking Theory for Altermagnetism in Bilayer Systems}},\ }\href
	{https://doi.org/10.1103/PhysRevLett.133.166701} {\bibfield  {journal}
		{\bibinfo  {journal} {Phys. Rev. Lett.}\ }\textbf {\bibinfo {volume} {133}},\
		\bibinfo {pages} {166701} (\bibinfo {year} {2024})}\BibitemShut {NoStop}%
	\bibitem [{\citenamefont {Qi}\ \emph {et~al.}(2024)\citenamefont {Qi},
		\citenamefont {Zhao},\ and\ \citenamefont {Zeng}}]{15}%
	\BibitemOpen
	\bibfield  {author} {\bibinfo {author} {\bibfnamefont {Y.}~\bibnamefont
			{Qi}}, \bibinfo {author} {\bibfnamefont {J.}~\bibnamefont {Zhao}},\ and\
		\bibinfo {author} {\bibfnamefont {H.}~\bibnamefont {Zeng}},\ }\bibfield
	{title} {\bibinfo {title} {{Spin-layer coupling in two-dimensional
				altermagnetic bilayers with tunable spin and valley splitting properties}},\
	}\href {https://doi.org/10.1103/PhysRevB.110.014442} {\bibfield  {journal}
		{\bibinfo  {journal} {Phys. Rev. B}\ }\textbf {\bibinfo {volume} {110}},\
		\bibinfo {pages} {014442} (\bibinfo {year} {2024})}\BibitemShut {NoStop}%
	\bibitem [{\citenamefont {Mavani}\ \emph {et~al.}(2025)\citenamefont {Mavani},
		\citenamefont {Huang}, \citenamefont {Samanta},\ and\ \citenamefont
		{Tsymbal}}]{16}%
	\BibitemOpen
	\bibfield  {author} {\bibinfo {author} {\bibfnamefont {H.}~\bibnamefont
			{Mavani}}, \bibinfo {author} {\bibfnamefont {K.}~\bibnamefont {Huang}},
		\bibinfo {author} {\bibfnamefont {K.}~\bibnamefont {Samanta}},\ and\ \bibinfo
		{author} {\bibfnamefont {E.~Y.}\ \bibnamefont {Tsymbal}},\ }\bibfield
	{title} {\bibinfo {title} {{Two-dimensional antiferromagnets with
				nonrelativistic spin splitting switchable by electric polarization}},\ }\href
	{https://doi.org/10.1103/c4dm-12ng} {\bibfield  {journal} {\bibinfo
			{journal} {Phys. Rev. B}\ }\textbf {\bibinfo {volume} {112}},\ \bibinfo
		{pages} {L060401} (\bibinfo {year} {2025})}\BibitemShut {NoStop}%
	\bibitem [{\citenamefont {Zeng}\ \emph {et~al.}(2026)\citenamefont {Zeng},
		\citenamefont {Liu}, \citenamefont {Peng}, \citenamefont {He}, \citenamefont
		{Yang},\ and\ \citenamefont {Zhao}}]{48-fop}%
	\BibitemOpen
	\bibfield  {author} {\bibinfo {author} {\bibfnamefont {S.}~\bibnamefont
			{Zeng}}, \bibinfo {author} {\bibfnamefont {D.}~\bibnamefont {Liu}}, \bibinfo
		{author} {\bibfnamefont {H.}~\bibnamefont {Peng}}, \bibinfo {author}
		{\bibfnamefont {C.-C.}\ \bibnamefont {He}}, \bibinfo {author} {\bibfnamefont
			{X.-B.}\ \bibnamefont {Yang}},\ and\ \bibinfo {author} {\bibfnamefont
			{Y.-J.}\ \bibnamefont {Zhao}},\ }\bibfield  {title} {\bibinfo {title}
		{{Classification and design of two-dimensional altermagnets}},\ }\href
	{https://doi.org/https://doi.org/10.15302/frontphys.2026.095301} {\bibfield
		{journal} {\bibinfo  {journal} {Front. Phys.}\ }\textbf {\bibinfo {volume}
			{21}},\ \bibinfo {pages} {095301} (\bibinfo {year} {2026})}\BibitemShut
	{NoStop}%
	\bibitem [{\citenamefont {Xu}\ \emph {et~al.}(2026)\citenamefont {Xu},
		\citenamefont {Gao},\ and\ \citenamefont {Liu}}]{17}%
	\BibitemOpen
	\bibfield  {author} {\bibinfo {author} {\bibfnamefont {R.}~\bibnamefont
			{Xu}}, \bibinfo {author} {\bibfnamefont {Y.}~\bibnamefont {Gao}},\ and\
		\bibinfo {author} {\bibfnamefont {J.}~\bibnamefont {Liu}},\ }\bibfield
	{title} {\bibinfo {title} {{Chemical design of monolayer altermagnets}},\
	}\href {https://doi.org/10.1093/nsr/nwaf528} {\bibfield  {journal} {\bibinfo
			{journal} {Natl. Sci. Rev.}\ }\textbf {\bibinfo {volume} {13}},\ \bibinfo
		{pages} {nwaf528} (\bibinfo {year} {2026})}\BibitemShut {NoStop}%
	\bibitem [{\citenamefont {Ma}\ \emph {et~al.}(2021)\citenamefont {Ma},
		\citenamefont {Hu}, \citenamefont {Li}, \citenamefont {Liu}, \citenamefont
		{Yao}, \citenamefont {Jia},\ and\ \citenamefont {Liu}}]{18}%
	\BibitemOpen
	\bibfield  {author} {\bibinfo {author} {\bibfnamefont {H.-Y.}\ \bibnamefont
			{Ma}}, \bibinfo {author} {\bibfnamefont {M.}~\bibnamefont {Hu}}, \bibinfo
		{author} {\bibfnamefont {N.}~\bibnamefont {Li}}, \bibinfo {author}
		{\bibfnamefont {J.}~\bibnamefont {Liu}}, \bibinfo {author} {\bibfnamefont
			{W.}~\bibnamefont {Yao}}, \bibinfo {author} {\bibfnamefont {J.-F.}\
			\bibnamefont {Jia}},\ and\ \bibinfo {author} {\bibfnamefont {J.}~\bibnamefont
			{Liu}},\ }\bibfield  {title} {\bibinfo {title} {{Multifunctional
				antiferromagnetic materials with giant piezomagnetism and noncollinear spin
				current}},\ }\href {https://doi.org/10.1038/s41467-021-23127-7} {\bibfield
		{journal} {\bibinfo  {journal} {Nat. Commun.}\ }\textbf {\bibinfo {volume}
			{12}},\ \bibinfo {pages} {2846} (\bibinfo {year} {2021})}\BibitemShut
	{NoStop}%
	\bibitem [{\citenamefont {Chen}\ \emph {et~al.}(2023)\citenamefont {Chen},
		\citenamefont {Wang}, \citenamefont {Li},\ and\ \citenamefont {Sanyal}}]{19}%
	\BibitemOpen
	\bibfield  {author} {\bibinfo {author} {\bibfnamefont {X.}~\bibnamefont
			{Chen}}, \bibinfo {author} {\bibfnamefont {D.}~\bibnamefont {Wang}}, \bibinfo
		{author} {\bibfnamefont {L.}~\bibnamefont {Li}},\ and\ \bibinfo {author}
		{\bibfnamefont {B.}~\bibnamefont {Sanyal}},\ }\bibfield  {title} {\bibinfo
		{title} {{Giant spin-splitting and tunable spin-momentum locked transport in
				room temperature collinear antiferromagnetic semimetallic CrO monolayer}},\
	}\href {https://doi.org/10.1063/5.0147450} {\bibfield  {journal} {\bibinfo
			{journal} {Appl. Phys. Lett.}\ }\textbf {\bibinfo {volume} {123}},\ \bibinfo
		{pages} {022402} (\bibinfo {year} {2023})}\BibitemShut {NoStop}%
	\bibitem [{\citenamefont {Che}\ \emph {et~al.}(2024)\citenamefont {Che},
		\citenamefont {Lv}, \citenamefont {Wu},\ and\ \citenamefont {Yang}}]{20}%
	\BibitemOpen
	\bibfield  {author} {\bibinfo {author} {\bibfnamefont {Y.}~\bibnamefont
			{Che}}, \bibinfo {author} {\bibfnamefont {H.}~\bibnamefont {Lv}}, \bibinfo
		{author} {\bibfnamefont {X.}~\bibnamefont {Wu}},\ and\ \bibinfo {author}
		{\bibfnamefont {J.}~\bibnamefont {Yang}},\ }\bibfield  {title} {\bibinfo
		{title} {Realizing altermagnetism in two-dimensional metal–organic
			framework semiconductors with electric-field-controlled anisotropic spin
			current},\ }\href {https://doi.org/10.1039/D4SC04125A} {\bibfield  {journal}
		{\bibinfo  {journal} {Chem. Sci.}\ }\textbf {\bibinfo {volume} {15}},\
		\bibinfo {pages} {13853} (\bibinfo {year} {2024})}\BibitemShut {NoStop}%
	\bibitem [{\citenamefont {Che}\ \emph {et~al.}(2025{\natexlab{a}})\citenamefont
		{Che}, \citenamefont {Lv}, \citenamefont {Wu},\ and\ \citenamefont
		{Yang}}]{21}%
	\BibitemOpen
	\bibfield  {author} {\bibinfo {author} {\bibfnamefont {Y.}~\bibnamefont
			{Che}}, \bibinfo {author} {\bibfnamefont {H.}~\bibnamefont {Lv}}, \bibinfo
		{author} {\bibfnamefont {X.}~\bibnamefont {Wu}},\ and\ \bibinfo {author}
		{\bibfnamefont {J.}~\bibnamefont {Yang}},\ }\bibfield  {title} {\bibinfo
		{title} {{Engineering Altermagnetic States in Two-Dimensional Square
				Tessellations}},\ }\href {https://doi.org/10.1103/v38b-5by1} {\bibfield
		{journal} {\bibinfo  {journal} {Phys. Rev. Lett.}\ }\textbf {\bibinfo
			{volume} {135}},\ \bibinfo {pages} {036701} (\bibinfo {year}
		{2025}{\natexlab{a}})}\BibitemShut {NoStop}%
	\bibitem [{\citenamefont {Peng}\ \emph {et~al.}(2025)\citenamefont {Peng},
		\citenamefont {Zeng}, \citenamefont {Liao}, \citenamefont {He}, \citenamefont
		{Yang},\ and\ \citenamefont {Zhao}}]{22}%
	\BibitemOpen
	\bibfield  {author} {\bibinfo {author} {\bibfnamefont {H.}~\bibnamefont
			{Peng}}, \bibinfo {author} {\bibfnamefont {S.}~\bibnamefont {Zeng}}, \bibinfo
		{author} {\bibfnamefont {J.-H.}\ \bibnamefont {Liao}}, \bibinfo {author}
		{\bibfnamefont {C.-C.}\ \bibnamefont {He}}, \bibinfo {author} {\bibfnamefont
			{X.-B.}\ \bibnamefont {Yang}},\ and\ \bibinfo {author} {\bibfnamefont
			{Y.-J.}\ \bibnamefont {Zhao}},\ }\bibfield  {title} {\bibinfo {title}
		{{Multicomponent altermagnet: A general approach to generating multicomponent
				structures with two-dimensional altermagnetism}},\ }\href
	{https://doi.org/10.1103/PhysRevB.111.195123} {\bibfield  {journal} {\bibinfo
			{journal} {Phys. Rev. B}\ }\textbf {\bibinfo {volume} {111}},\ \bibinfo
		{pages} {195123} (\bibinfo {year} {2025})}\BibitemShut {NoStop}%
	\bibitem [{\citenamefont {Mazin}\ \emph {et~al.}(2023)\citenamefont {Mazin},
		\citenamefont {Gonz{\'a}lez-Hern{\'a}ndez},\ and\ \citenamefont
		{{\v{S}}mejkal}}]{23}%
	\BibitemOpen
	\bibfield  {author} {\bibinfo {author} {\bibfnamefont {I.}~\bibnamefont
			{Mazin}}, \bibinfo {author} {\bibfnamefont {R.}~\bibnamefont
			{Gonz{\'a}lez-Hern{\'a}ndez}},\ and\ \bibinfo {author} {\bibfnamefont
			{L.}~\bibnamefont {{\v{S}}mejkal}},\ }\bibfield  {title} {\bibinfo {title}
		{{Induced Monolayer Altermagnetism in MnP(S,Se)$_3$ and FeSe}},\ }\href
	{https://arxiv.org/abs/2309.02355} {\bibfield  {journal} {\bibinfo  {journal}
			{arXiv}\ } (\bibinfo {year} {2023})}\BibitemShut {NoStop}%
	\bibitem [{\citenamefont {Ma}\ \emph {et~al.}(2025)\citenamefont {Ma},
		\citenamefont {Zhang}, \citenamefont {Gao}, \citenamefont {Wan},
		\citenamefont {Ma}, \citenamefont {Jiao},\ and\ \citenamefont {Yang}}]{24}%
	\BibitemOpen
	\bibfield  {author} {\bibinfo {author} {\bibfnamefont {F.}~\bibnamefont
			{Ma}}, \bibinfo {author} {\bibfnamefont {Z.}~\bibnamefont {Zhang}}, \bibinfo
		{author} {\bibfnamefont {Z.}~\bibnamefont {Gao}}, \bibinfo {author}
		{\bibfnamefont {X.}~\bibnamefont {Wan}}, \bibinfo {author} {\bibfnamefont
			{Y.}~\bibnamefont {Ma}}, \bibinfo {author} {\bibfnamefont {Y.}~\bibnamefont
			{Jiao}},\ and\ \bibinfo {author} {\bibfnamefont {S.~A.}\ \bibnamefont
			{Yang}},\ }\bibfield  {title} {\bibinfo {title} {{Engineering Spin Splitting
				in Antiferromagnets by Superatoms with Internal Degrees of Freedom}},\ }\href
	{https://doi.org/10.1021/acs.nanolett.5c04171} {\bibfield  {journal}
		{\bibinfo  {journal} {Nano Lett.}\ }\textbf {\bibinfo {volume} {25}},\
		\bibinfo {pages} {15377} (\bibinfo {year} {2025})}\BibitemShut {NoStop}%
	\bibitem [{\citenamefont {Wang}\ \emph {et~al.}(2023)\citenamefont {Wang},
		\citenamefont {Impundu}, \citenamefont {Jin}, \citenamefont {Peng},
		\citenamefont {Liu}, \citenamefont {Wei}, \citenamefont {Xu}, \citenamefont
		{Wang}, \citenamefont {You}, \citenamefont {Fan}, \citenamefont {Li},\ and\
		\citenamefont {Sun}}]{25}%
	\BibitemOpen
	\bibfield  {author} {\bibinfo {author} {\bibfnamefont {W.}~\bibnamefont
			{Wang}}, \bibinfo {author} {\bibfnamefont {J.}~\bibnamefont {Impundu}},
		\bibinfo {author} {\bibfnamefont {J.}~\bibnamefont {Jin}}, \bibinfo {author}
		{\bibfnamefont {Z.}~\bibnamefont {Peng}}, \bibinfo {author} {\bibfnamefont
			{H.}~\bibnamefont {Liu}}, \bibinfo {author} {\bibfnamefont {Z.}~\bibnamefont
			{Wei}}, \bibinfo {author} {\bibfnamefont {Y.}~\bibnamefont {Xu}}, \bibinfo
		{author} {\bibfnamefont {Y.}~\bibnamefont {Wang}}, \bibinfo {author}
		{\bibfnamefont {J.}~\bibnamefont {You}}, \bibinfo {author} {\bibfnamefont
			{W.}~\bibnamefont {Fan}}, \bibinfo {author} {\bibfnamefont {Y.~J.}\
			\bibnamefont {Li}},\ and\ \bibinfo {author} {\bibfnamefont {L.}~\bibnamefont
			{Sun}},\ }\bibfield  {title} {\bibinfo {title} {{Ferromagnetism in sp$^2$
				carbon}},\ }\href {https://doi.org/10.1007/s12274-023-5972-8} {\bibfield
		{journal} {\bibinfo  {journal} {Nano Res.}\ }\textbf {\bibinfo {volume}
			{16}},\ \bibinfo {pages} {12883} (\bibinfo {year} {2023})}\BibitemShut
	{NoStop}%
	\bibitem [{\citenamefont {Liao}\ \emph {et~al.}(2025)\citenamefont {Liao},
		\citenamefont {Wang}, \citenamefont {Ye}, \citenamefont {Wu}, \citenamefont
		{Jiang}, \citenamefont {Zhou}, \citenamefont {Guan},\ and\ \citenamefont
		{Wang}}]{26}%
	\BibitemOpen
	\bibfield  {author} {\bibinfo {author} {\bibfnamefont {M.}~\bibnamefont
			{Liao}}, \bibinfo {author} {\bibfnamefont {Y.}~\bibnamefont {Wang}}, \bibinfo
		{author} {\bibfnamefont {P.}~\bibnamefont {Ye}}, \bibinfo {author}
		{\bibfnamefont {C.}~\bibnamefont {Wu}}, \bibinfo {author} {\bibfnamefont
			{H.}~\bibnamefont {Jiang}}, \bibinfo {author} {\bibfnamefont
			{F.}~\bibnamefont {Zhou}}, \bibinfo {author} {\bibfnamefont {J.}~\bibnamefont
			{Guan}},\ and\ \bibinfo {author} {\bibfnamefont {F.}~\bibnamefont {Wang}},\
	}\bibfield  {title} {\bibinfo {title} {{AM-C33}: an altermagnetic carbon},\
	}\href {https://doi.org/10.1039/D5SC05194K} {\bibfield  {journal} {\bibinfo
			{journal} {Chem. Sci.}\ }\textbf {\bibinfo {volume} {16}},\ \bibinfo {pages}
		{23174} (\bibinfo {year} {2025})}\BibitemShut {NoStop}%
	\bibitem [{\citenamefont {Vi{\~n}a-Baus{\'a}}\ \emph
		{et~al.}(2025)\citenamefont {Vi{\~n}a-Baus{\'a}}, \citenamefont
		{Garc{\'i}a-Bl{\'a}zquez}, \citenamefont {Chourasia}, \citenamefont
		{Carrasco}, \citenamefont {Exp{\'o}sito}, \citenamefont {Brihuega},\ and\
		\citenamefont {Palacios}}]{27}%
	\BibitemOpen
	\bibfield  {author} {\bibinfo {author} {\bibfnamefont {B.}~\bibnamefont
			{Vi{\~n}a-Baus{\'a}}}, \bibinfo {author} {\bibfnamefont {M.~A.}\ \bibnamefont
			{Garc{\'i}a-Bl{\'a}zquez}}, \bibinfo {author} {\bibfnamefont
			{S.}~\bibnamefont {Chourasia}}, \bibinfo {author} {\bibfnamefont
			{R.}~\bibnamefont {Carrasco}}, \bibinfo {author} {\bibfnamefont
			{D.}~\bibnamefont {Exp{\'o}sito}}, \bibinfo {author} {\bibfnamefont
			{I.}~\bibnamefont {Brihuega}},\ and\ \bibinfo {author} {\bibfnamefont
			{J.~J.}\ \bibnamefont {Palacios}},\ }\bibfield  {title} {\bibinfo {title}
		{{Building Unconventional Magnetic Phases on Graphene by H Atom Manipulation:
				From Altermagnets to Lieb Ferrimagnets}},\ }\href
	{https://doi.org/10.1021/acs.nanolett.5c02091} {\bibfield  {journal}
		{\bibinfo  {journal} {Nano Lett.}\ }\textbf {\bibinfo {volume} {25}},\
		\bibinfo {pages} {11554} (\bibinfo {year} {2025})}\BibitemShut {NoStop}%
	\bibitem [{\citenamefont {Li}\ and\ \citenamefont {Shao}(2025)}]{28}%
	\BibitemOpen
	\bibfield  {author} {\bibinfo {author} {\bibfnamefont {H.-B.}\ \bibnamefont
			{Li}}\ and\ \bibinfo {author} {\bibfnamefont {Z.-G.}\ \bibnamefont {Shao}},\
	}\bibfield  {title} {\bibinfo {title} {{Achieving altermagnetism in monolayer
				holey graphyne via atomic manipulation}},\ }\href
	{https://doi.org/10.1103/72r8-lq5r} {\bibfield  {journal} {\bibinfo
			{journal} {Phys. Rev. B}\ }\textbf {\bibinfo {volume} {111}},\ \bibinfo
		{pages} {214420} (\bibinfo {year} {2025})}\BibitemShut {NoStop}%
	\bibitem [{\citenamefont {Guo}\ \emph {et~al.}(2025)\citenamefont {Guo},
		\citenamefont {Zong}, \citenamefont {Zhang}, \citenamefont {Wang},
		\citenamefont {Liu},\ and\ \citenamefont {Ji}}]{38}%
	\BibitemOpen
	\bibfield  {author} {\bibinfo {author} {\bibfnamefont {D.}~\bibnamefont
			{Guo}}, \bibinfo {author} {\bibfnamefont {C.}~\bibnamefont {Zong}}, \bibinfo
		{author} {\bibfnamefont {W.}~\bibnamefont {Zhang}}, \bibinfo {author}
		{\bibfnamefont {C.}~\bibnamefont {Wang}}, \bibinfo {author} {\bibfnamefont
			{J.}~\bibnamefont {Liu}},\ and\ \bibinfo {author} {\bibfnamefont
			{W.}~\bibnamefont {Ji}},\ }\bibfield  {title} {\bibinfo {title} {{Tunable
				altermagnetism via interchain engineering in parallel-assembled atomic
				chains}},\ }\href {https://doi.org/10.1103/zhds-vnyt} {\bibfield  {journal}
		{\bibinfo  {journal} {Phys. Rev. B}\ }\textbf {\bibinfo {volume} {112}},\
		\bibinfo {pages} {L041404} (\bibinfo {year} {2025})}\BibitemShut {NoStop}%
	\bibitem [{\citenamefont {C{\^o}t{\'e}}\ \emph {et~al.}(2005)\citenamefont
		{C{\^o}t{\'e}}, \citenamefont {Benin}, \citenamefont {Ockwig}, \citenamefont
		{O'Keeffe}, \citenamefont {Matzger},\ and\ \citenamefont {Yaghi}}]{39}%
	\BibitemOpen
	\bibfield  {author} {\bibinfo {author} {\bibfnamefont {A.~P.}\ \bibnamefont
			{C{\^o}t{\'e}}}, \bibinfo {author} {\bibfnamefont {A.~I.}\ \bibnamefont
			{Benin}}, \bibinfo {author} {\bibfnamefont {N.~W.}\ \bibnamefont {Ockwig}},
		\bibinfo {author} {\bibfnamefont {M.}~\bibnamefont {O'Keeffe}}, \bibinfo
		{author} {\bibfnamefont {A.~J.}\ \bibnamefont {Matzger}},\ and\ \bibinfo
		{author} {\bibfnamefont {O.~M.}\ \bibnamefont {Yaghi}},\ }\bibfield  {title}
	{\bibinfo {title} {{Porous, Crystalline, Covalent Organic Frameworks}},\
	}\href {https://doi.org/10.1126/science.1120411} {\bibfield  {journal}
		{\bibinfo  {journal} {Science}\ }\textbf {\bibinfo {volume} {310}},\ \bibinfo
		{pages} {1166} (\bibinfo {year} {2005})}\BibitemShut {NoStop}%
	\bibitem [{\citenamefont {Du}\ \emph {et~al.}(2023)\citenamefont {Du},
		\citenamefont {Gao}, \citenamefont {Xu},\ and\ \citenamefont {Xu}}]{47}%
	\BibitemOpen
	\bibfield  {author} {\bibinfo {author} {\bibfnamefont {L.}~\bibnamefont
			{Du}}, \bibinfo {author} {\bibfnamefont {B.}~\bibnamefont {Gao}}, \bibinfo
		{author} {\bibfnamefont {S.}~\bibnamefont {Xu}},\ and\ \bibinfo {author}
		{\bibfnamefont {Q.}~\bibnamefont {Xu}},\ }\bibfield  {title} {\bibinfo
		{title} {{Strong ferromagnetism of g-C$_{3}$N$_{4}$ achieved by atomic
				manipulation}},\ }\href {https://doi.org/10.1038/s41467-023-38012-8}
	{\bibfield  {journal} {\bibinfo  {journal} {Nat. Commun.}\ }\textbf {\bibinfo
			{volume} {14}},\ \bibinfo {pages} {2278} (\bibinfo {year}
		{2023})}\BibitemShut {NoStop}%
	\bibitem [{\citenamefont {Che}\ \emph {et~al.}(2025{\natexlab{b}})\citenamefont
		{Che}, \citenamefont {Chen}, \citenamefont {Liu}, \citenamefont {Lv},
		\citenamefont {Wu},\ and\ \citenamefont {Yang}}]{41}%
	\BibitemOpen
	\bibfield  {author} {\bibinfo {author} {\bibfnamefont {Y.}~\bibnamefont
			{Che}}, \bibinfo {author} {\bibfnamefont {Y.}~\bibnamefont {Chen}}, \bibinfo
		{author} {\bibfnamefont {X.}~\bibnamefont {Liu}}, \bibinfo {author}
		{\bibfnamefont {H.}~\bibnamefont {Lv}}, \bibinfo {author} {\bibfnamefont
			{X.}~\bibnamefont {Wu}},\ and\ \bibinfo {author} {\bibfnamefont
			{J.}~\bibnamefont {Yang}},\ }\bibfield  {title} {\bibinfo {title} {{Inverse
				Design of 2D Altermagnetic Metal--Organic Framework Monolayers from
				H{\"u}ckel Theory of Nonbonding Molecular Orbitals}},\ }\href
	{https://doi.org/10.1021/jacsau.4c01150} {\bibfield  {journal} {\bibinfo
			{journal} {JACS Au}\ }\textbf {\bibinfo {volume} {5}},\ \bibinfo {pages}
		{381} (\bibinfo {year} {2025}{\natexlab{b}})}\BibitemShut {NoStop}%
	\bibitem [{\citenamefont {Goodenough}(1955)}]{42}%
	\BibitemOpen
	\bibfield  {author} {\bibinfo {author} {\bibfnamefont {J.~B.}\ \bibnamefont
			{Goodenough}},\ }\bibfield  {title} {\bibinfo {title} {{Theory of the Role of
				Covalence in the Perovskite-Type Manganites $[\mathrm{La},
				M(\mathrm{II})]\mathrm{Mn}{\mathrm{O}}_{3}$}},\ }\href
	{https://doi.org/10.1103/PhysRev.100.564} {\bibfield  {journal} {\bibinfo
			{journal} {Phys. Rev.}\ }\textbf {\bibinfo {volume} {100}},\ \bibinfo {pages}
		{564} (\bibinfo {year} {1955})}\BibitemShut {NoStop}%
	\bibitem [{\citenamefont {Kanamori}(1959)}]{43}%
	\BibitemOpen
	\bibfield  {author} {\bibinfo {author} {\bibfnamefont {J.}~\bibnamefont
			{Kanamori}},\ }\bibfield  {title} {\bibinfo {title} {{Superexchange
				interaction and symmetry properties of electron orbitals}},\ }\href
	{https://doi.org/https://doi.org/10.1016/0022-3697(59)90061-7} {\bibfield
		{journal} {\bibinfo  {journal} {J. Phys. Chem. Solids}\ }\textbf {\bibinfo
			{volume} {10}},\ \bibinfo {pages} {87} (\bibinfo {year} {1959})}\BibitemShut
	{NoStop}%
	\bibitem [{\citenamefont {Anderson}(1950)}]{44}%
	\BibitemOpen
	\bibfield  {author} {\bibinfo {author} {\bibfnamefont {P.~W.}\ \bibnamefont
			{Anderson}},\ }\bibfield  {title} {\bibinfo {title} {{Antiferromagnetism.
				Theory of Superexchange Interaction}},\ }\href
	{https://doi.org/10.1103/PhysRev.79.350} {\bibfield  {journal} {\bibinfo
			{journal} {Phys. Rev.}\ }\textbf {\bibinfo {volume} {79}},\ \bibinfo {pages}
		{350} (\bibinfo {year} {1950})}\BibitemShut {NoStop}%
	\bibitem [{\citenamefont {Lu}\ \emph {et~al.}(2021)\citenamefont {Lu},
		\citenamefont {Jiao}, \citenamefont {Li}, \citenamefont {Yi},\ and\
		\citenamefont {Zhang}}]{45}%
	\BibitemOpen
	\bibfield  {author} {\bibinfo {author} {\bibfnamefont {H.-Y.}\ \bibnamefont
			{Lu}}, \bibinfo {author} {\bibfnamefont {N.}~\bibnamefont {Jiao}}, \bibinfo
		{author} {\bibfnamefont {B.-W.}\ \bibnamefont {Li}}, \bibinfo {author}
		{\bibfnamefont {W.-C.}\ \bibnamefont {Yi}},\ and\ \bibinfo {author}
		{\bibfnamefont {P.}~\bibnamefont {Zhang}},\ }\bibfield  {title} {\bibinfo
		{title} {{Hydrogenated group IV-V monolayer HAB$_6$: A new type of Dirac
				material constructed by isoelectronic rule}},\ }\href
	{https://doi.org/https://doi.org/10.1016/j.apsusc.2021.149635} {\bibfield
		{journal} {\bibinfo  {journal} {Appl. Surf. Sci.}\ }\textbf {\bibinfo
			{volume} {554}},\ \bibinfo {pages} {149635} (\bibinfo {year}
		{2021})}\BibitemShut {NoStop}%
	\bibitem [{\citenamefont {Yi}\ \emph {et~al.}(2021)\citenamefont {Yi},
		\citenamefont {Jiang}, \citenamefont {Wang}, \citenamefont {Yang},
		\citenamefont {Yang},\ and\ \citenamefont {Liu}}]{46}%
	\BibitemOpen
	\bibfield  {author} {\bibinfo {author} {\bibfnamefont {W.}~\bibnamefont
			{Yi}}, \bibinfo {author} {\bibfnamefont {X.}~\bibnamefont {Jiang}}, \bibinfo
		{author} {\bibfnamefont {Z.}~\bibnamefont {Wang}}, \bibinfo {author}
		{\bibfnamefont {T.}~\bibnamefont {Yang}}, \bibinfo {author} {\bibfnamefont
			{B.}~\bibnamefont {Yang}},\ and\ \bibinfo {author} {\bibfnamefont
			{X.}~\bibnamefont {Liu}},\ }\bibfield  {title} {\bibinfo {title} {{ABX$_6$
				Monolayers: A new Dirac material family containing high Fermi velocities and
				topological properties}},\ }\href
	{https://doi.org/https://doi.org/10.1016/j.apsusc.2021.151237} {\bibfield
		{journal} {\bibinfo  {journal} {Appl. Surf. Sci.}\ }\textbf {\bibinfo
			{volume} {570}},\ \bibinfo {pages} {151237} (\bibinfo {year}
		{2021})}\BibitemShut {NoStop}%
\end{thebibliography}
\end{document}